\documentstyle[psfig,11pt]{article}
\title{\sc X-ray temperature distribution function of clusters in presence of 
large non-radial motions} 
\author{{\bf A. Del Popolo} ~ and ~ {\bf M. Gambera}}
\date{}
\begin{document}
\maketitle
\begin{center}
{\sc Istituto di Astronomia dell'Universit\`a di Catania,\\ 
Viale A.Doria, 6 - I 95125 Catania, ITALY \\}
~\\

{\bf To be published:}\\
In the Proceedings of {\it " The VII Conference on Theoretical Physics: 
General Relativity and Gravitation "} \\
BISTRITZA - MAY 26-30, 1997 - Romania\\
\end{center}
~\\
\begin{abstract}
We consider the non-radial motions, originating
in the outskirts of clusters of galaxies and we show how it may
reduce the discrepancy between Cold Dark Matter (CDM) predicted X-ray 
temperature distribution function of clusters of galaxies and that 
really observed, so resolving the problem of the X-ray clusters 
abundance over-production predicted by the CDM model. We construct
the X-ray temperature distribution function using Press-Schechter's 
(1974) theory and Evrard's (1990) prescriptions for the mass-temperature 
relation, taking also account of the non-radial motions
originating from the gravitational interaction of the quadrupole moment
of the protocluster. Then, we compare our X-ray temperature distribution 
function with the tidal field of the matter of the neighboring 
protostructures. We find that the model produces a reasonable 
cluster temperature distribution.\\
~\\
keywords: ~ {\it cosmology: theory-large scale structure of Universe - galaxies: formation}

\end{abstract}
\newpage

\section{Introduction}

In the last years the CDM model has contributed to obtain a better
understanding of the origin and evolution of the large scale structure in the
Universe (White et al. 1987; Frenk et al. 1988; Efstathiou 1990; Ostriker
1993). The principal assumptions of the SCDM (standard CDM) model are: 
{\bf a)} a flat Universe dominated by weakly
interacting elementary particles having low velocity dispersion at early
times; {\bf b)} Barionic content determined by the standard big bang
nucleosynthesis model (Kernan \& Sarkar 1996; Steigman 1996; Olive 1997; 
Dolgov 1997); 
{\bf c)} critical matter density; 
{\bf d)} expansion rate given by $ h = 0.5$; 
{\bf e)} a scale invariant and adiabatic spectrum with a spectral index, 
$ n \equiv 1$; 
{\bf f)} the condition required by observations, that the fluctuations
in galaxy distribution, $(\delta \rho / \rho)_{g}$ , are larger than the
fluctuations in the mass distribution, $(\delta \rho / \rho)_{\rho} $ by 
a factor $ b > 1$. 
As pointed out by Davis et al. (1985), without this last assumption, 
a unbiased CDM (that is $ b = 0$) presents of several problems: 
pairwise velocity dispersion larger than the observed one, galaxy 
correlation function steeper than observed (Liddle \& Lyth 1993 and 
Strauss \& Willick 1995). 
The remedy to these problems is the concept of biasing (Kaiser 1984), 
i.e. that galaxies are more strongly clustered than the mass distribution 
from which they originated.
The physical origin of such biasing is not yet clear even if several
mechanisms have been proposed (Rees 1985; Dekel \& Silk 1986; Dekel \& 
Rees 1987; Carlberg 1991; Cen \& Ostriker 1992; Bower et al. 1993; Silk \& 
Wyse 1993).\\
Although at his appearance the standard form
of CDM was very successful in describing the observed structures in the
Universe (galaxy clustering statistics, structure formation epochs, peculiar
velocity flows) (Peebles, 1982; Blumenthal et al. 1984; Bardeen et al. 1986;
White et al. 1987; Frenk et al. 1988; Efstathiou 1990) recent measurements
have shown several deficiencies of the model, at least if any bias of the
distribution of galaxies relative to the mass is constant with scale. Some of 
the most difficult problems to reconcile with the theory  are the magnitude of the dipole of the angular distribution of optically selected galaxies (Lahav 
et al. 1988; Kaiser \& Lahav 1989), the possible observations of clusters of
galaxies with high velocity dispersion at $z\geq 0.5$ (Evrard 1989), the 
strong clustering of rich clusters of galaxies, 
$ \xi_{cc}(r) \simeq (r / 25h^{-1}Mpc)^{-2}$, 
far in excess of CDM predictions (Bahcall \&
Soneira 1983), the X-ray temperature distribution function of clusters,
over-producing the observed cluster abundances (Bartlett \& Silk 1993), the
conflict between the normalisation of the spectrum of the perturbation which
is required by different types of observations. Normalization obtained from
COBE data (Smoot et al. 1992) on scales of the order of $10^3Mpc$ 
requires $\sigma_8=0.95\pm 0.2$. 
Normalisation on scales $10$ to $50Mpc$ obtained from QDOT
(Kaiser et al. 1991) and POTENT (Dekel et al. 1992) requires that $\sigma _8$
is in the range $0.7\div 1.1$, which is compatible with COBE normalisation
while the observations of the pairwise velocity dispersion of galaxies on
scales $r\leq 3Mpc$ seem to require $\sigma _8<0.5$. Another problem of CDM
model is the incorrect scale dependence of the galaxy correlation
function, $\xi (r)$, on scales $10$ to $100$ $Mpc$, having $\xi (r)$ too
little power on the large scales compared to the power on smaller scales.
The APM survey (Maddox et al. 1990, 1991), giving the galaxy angular
correlation function, the 1.2 Jy IRAS power spectrum, the QDOT survey
(Saunders et al. 1991), X-ray observations (Lahav et al. 1989) and radio
observations (Peacock 1991; Peacock \& Nicholson 1991) agree with the quoted
conclusion. As shown in recent studies of galaxy clustering on large scales
(Maddox et al. 1990; Efstathiou et al. 1990b; Saunders et al. 1991) the
measured rms fluctuations within spheres of radius $20h^{-1}Mpc$ have
value 2-3 times larger than that predicted by the CDM model.\\
Alternative models with more large-scale power than CDM have been introduced
in order to solve the latter problem (Peebles 1984; Shafi \& Stecker 1984; 
Valdarnini \& Bonometto 1985; Bond et al. 1988; Holtzman 1989;
Efstathiou et al. 1990a; Schaefer 1991; Turner 1991; 
Holtzman \& Primack 1993; Shaefer \& Shafi 1993; White et al. 1993).\\
In this work, we show how the discrepancy between the predicted and 
observed X-ray temperature distribution function of clusters may be reduced 
when non-radial motions, that develop during the collapse process,
are taken into account. We compare the distribution function obtained 
taking into account non-radial motions with X-ray temperature 
distributions obteined using a simple CDM spectrum.\\
The plan of this paper is the following: in \S ~2 we find the effects 
of non-radial motions on the X-ray temperature distribution function and 
then we compare this prevision to that by a simple CDM model. In \S ~3 is 
devoted to the discussion of our results.\\

\section{X-ray temperature distribution function}

The Press-Schechter theory provides an analytical description of the
evolution of structures in a hierarchical Universe. In this model the linear
density field, $\rho (x,t)$, is an isotropic random Gaussian field, the
non-linear clumps are identified as over-densities (having a density contrast 
$\delta _c\sim 1.68$; Gunn \& Gott 1972) in the linear density field, while
a mass element is incorporated into a non-linear object of mass $M$ when the
density field smoothed with a top-hat filter of radius $R_f$, exceeds a
threshold $\delta _c$ ($ M\propto R_f^3$). The fraction of mass in objects of
mass greater than $ M$ is given by:
\begin{equation}
F(>M)=\frac 1{(2\pi )^{1/2}\sigma }\int_{\delta _c}^\infty \exp \left( \frac{%
-\delta _c^2}{2\sigma ^2}\right) d\delta 
\label{eq:ma1}
\end{equation}
while the comoving number density of non-linear objects of mass $M$ to $M+dM 
$ is given simply by the derivative of $F(>M)$:
\begin{equation}
N(M,t)dM=-\rho \sqrt{\frac 2\pi }\nu \exp \left( -\nu ^2/2\right) \frac
1\sigma \left( \frac{d\sigma }{dM}\right) \frac{dM}M  
\label{eq:press}
\end{equation}
where $\rho $ is the mean mass density, $\nu = \frac{\delta _c}{\sigma (M)}$ 
and $\sigma (M)$ is the linear power spectrum evaluated at the epoch when 
the mass function is desidered, given by 
\begin{equation}
\sigma ^2(M)=\frac 1{2\pi ^2}\int_0^\infty dkk^2P(k)W^2(kR) 
\label{eq:ma3}
\end{equation}
where $ P(k)$ is the power spectrum and $ W(kR)$ is a top-hat 
smoothing function:
\begin{equation}
W(kR)=\frac 3{\left( kR\right) ^3}\left( \sin kR-kR\cos kR\right) 
\label{eq:ma4}
\end{equation}
The redshift dependence of
Eq. ~(\ref{eq:press}) can be
obtained remembering that
\begin{equation}
\nu =\frac{\delta _c(z)D(0)}{\sigma _o(M)D(z)}
\end{equation}
being $ D(z)$ the growth factor of the density perturbation.
In Eq. ~(\ref{eq:press})
Press-Schechter introduced arbitrarily a factor of two because 
\begin{equation}
\int_0^\infty dF(M) = \frac{1}{2} 
\nonumber
\end{equation}
so that only half of the mass in the Universe
is accounted for. Several papers have tried to solve the problems of 
this model in order to improve it (Colafrancesco et al. 1989; 
Peacock \& Heavens 1990; Bower 1991). In spite of the quoted problems, 
N-body simulations developed by Efstathiou et al. (1988), Brainerd \& 
Villumsen (1992) showed that Press-Schechter analytic theory provides a good 
description of the evolution of the distribution of mass amongst groups and 
clusters of galaxies (multiplicity function) and this has proven particularly 
useful in analyzing the number counts and redshift distributions for QSOs 
(Efstathiou \& Rees 1988), Lyman $\alpha $ clouds (Bond et al. 1988) and 
X-ray clusters (Cavaliere \& Colafrancesco 1988).\\
The mass function can be transformed into the temperature distribution if 
a relation between the mass and the temperature is known. For a flat 
matter-dominated Universe a simple scaling gives (Kaiser 1986; Evrard 1990):
\begin{equation}
T=(6.4h^{2/3}keV)\left( \frac M{10^{15}M_{\odot }}\right) ^{2/3}(1+z) 
\label{eq:ma2}
\end{equation}
The mass variance present in Eq.~(\ref{eq:press})
can be obtained remembering the Eq.~(\ref{eq:ma3}) and once a power 
spectrum, $P(k)=Ak^nT^2(k)$ is fixed giving the transfer
function $T(k):$
\begin{eqnarray}
T(k) &=& [\ln \left( 1+4.164k\right)]^2 \cdot (192.9+1340k+ \nonumber \\
& + &  1.599\cdot 10^5k^2+1.78\cdot 10^5k^3+3.995\cdot
10^6k^4)^{-1/2}
\label{eq:ma5}
\end{eqnarray}
(Ryden \& Gunn 1987) and $ A$ is the normalizing constant. The normalization 
is obtained, as usual, imposing that the variance in a sphere of $8h^{-1}Mpc$, 
$\sigma _{8}$, is unity. As shown by Bartlett \& Silk (1993) the X-ray 
distribution function obtained using a standard CDM spectrum over-produces 
the clusters abundances. The discrepancy can be reduced taking into account 
the non-radial motions that originate when a cluster reaches the non-linear 
regime.
In fact, the Press-Schechter temperature distribution requires specification
of $\delta _c$ and the temperature-mass relation. The presence of non-radial
motions changes both $\delta _c$ and the T-M relation. As
shown by Barrow \& Silk (1981) and Szalay \& Silk (1983) the gravitational
interaction of the irregular mass distribution of proto-cluster with the
neighbouring proto-structures gives rise to non-radial motions, within the
protocluster, which are expected to slow the rate of growth of the 
density contrast and to delay or suppress the collapse. 
According to Davis \& Peebles (1977) the kinetic energy
of the resulting non-radial motions at the epoch of maximum expansion
increases so much to oppose the recollapse of the proto-structure. Numerical
N-body simulations by Villumsen \& Davis (1986) showed a tendency to
reproduce this so called previrialization effect. In a more recent paper by
Peebles (1990) the slowdown of the growth of density fluctuations and the
collapse suppression after the epoch of the maximum expansion were re-obtained
using a numerical action method.
In the central regions of a density peak ($r\leq 0.5R_f$) the 
velocity dispersion attain nearly the same value (Antonuccio-Delogu \& 
Colafrancesco 1995) while at larger radii ($r \geq R_f$) the radial 
component is lower than the tangential component. 
This means that motions in the outer regions are predominantly non-radial and 
in these regions the fate of the infalling material could be influenced by 
the amount of tangential velocity relative to the radial one. This can be 
shown writing the equation of motion of a spherically symmetric 
mass distribution with density $n(r)$:
\begin{equation}
\frac \partial {\partial t}n \langle v_r \rangle +\frac \partial {\partial r}n 
\langle v_r^2 \rangle + \left(2 \langle v_r^2 \rangle - 
\langle v_\vartheta ^2 \rangle \right) \frac nr+n(r)\frac \partial {\partial
t} \langle v_r \rangle = 0  
\label{eq:peeb}
\end{equation}
where $ \langle v_r \rangle$ and $ \langle v_\vartheta \rangle $ 
are, respectively, the mean radial and
tangential streaming velocity. Eq. (\ref{eq:peeb}) shows that high
tangential velocity dispersion 
$(\langle v_\vartheta ^2 \rangle \geq 2 \langle v_r^2 \rangle)$ 
may alter the infall pattern. The expected delay in the collapse of a 
perturbation may be calculated solving the equation for the radial 
acceleration (Peebles 1993): 
\begin{equation}
\frac{dv_r}{dt}=\frac{L^2(r,\nu )}{M^2r^3}-g(r)  
\label{eq:coll}
\end{equation}
where $L(r,\nu )$ is the angular momentum and $g(r)$ the acceleration.
Writing the proper radius of a shell in terms of the expansion parameter, $%
a(r_i,t)$: 
\begin{equation}
r(r_i,t)=r_{i}a(r_i,t)
\label{eq:ma6}
\end{equation}
remembering that $M=\frac{4\pi }3\rho _b(r_i,t)a^3(r_i,t)r_i^3$, that $\frac{%
3H^2}{8\pi G}=\rho _b$, where $H$ is the Hubble constant, and assuming that
no shell crossing occurs so that the total mass inside each shell remains
constant, ($\rho (r_i,t)=\frac{\rho _i(r_i,t)}{a^3(r_i,t)}$) Eq. (\ref
{eq:coll}) may be written as: 
\begin{equation}
\frac{d^2a}{dt^2}=-\frac{H^2(1+\overline{\delta })}{2a^2}+\frac{4G^2L^2}{%
H^4(1+\overline{\delta })^2r_i^{10}a^3}  
\label{eq:sec}
\end{equation}
or integrating the equation once more: 
\begin{equation}
(\frac{da}{dt})^2=H_i^2\left[ \frac{1+\overline{\delta }}a\right] +\int 
\frac{8G^2L^2}{H_i^2r_i^{10}\left( 1+\overline{\delta }\right) ^2}\frac
1{a^3}da-2C  
\label{eq:ses}
\end{equation}
where $\overline{\delta }=\frac{\rho _i-\rho _0}{\rho _0}$ is 
the overdensity and $ C$ is the binding energy of the shell. 

\begin{figure}
\psfig{file=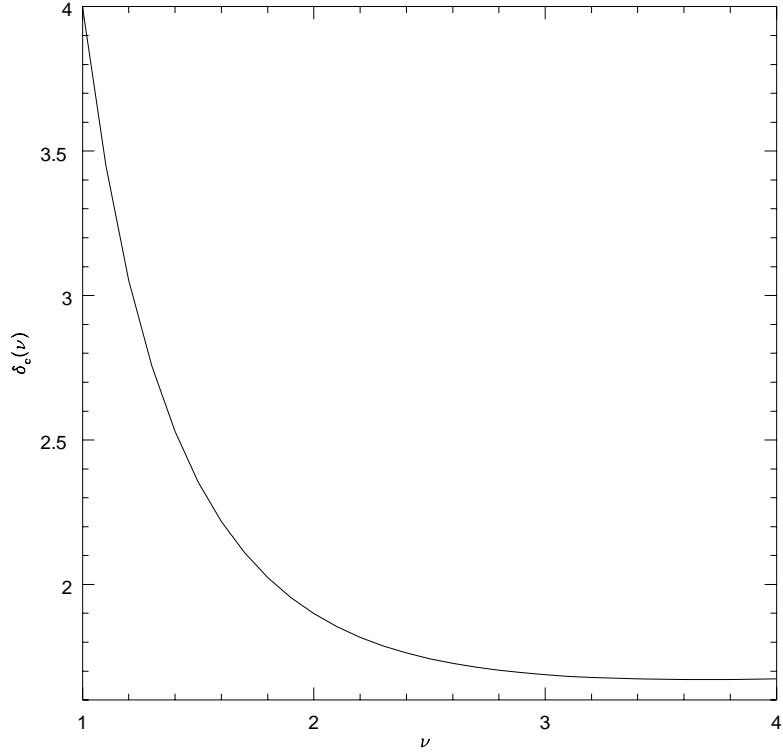,width=11cm}
\vspace*{2.0 cm}
\caption[]{The threshold $\delta_{c}$ in function of the mass M, for a CDM
spectrum with $R_{f}=3 Mpc$, taking account of non-radial motions.}
\end{figure}

\newpage

Using a technique similar to that by Bartlett \& Silk (1993) 
it is possible to obtain the overdensity at the time of collapse:
\begin{equation}
\delta _c(\nu )=\delta _{co}\left[ 1+\frac{8G^2}{\Omega
_o^3H_o^6r_i^{10}\overline{\delta} (1+\overline{\delta} )^2}\int_0^{a_{\max }}\frac{L^2 \cdot da}{a^3}%
\right]
\label{eq:ma7} 
\end{equation}
where $\delta _{co}=1.68$ is the critical threshold for a spherical model.
Filtering the spectrum on cluster scales, $R_f=3h^{-1}Mpc$, we obtained the
total specific angular momentum, $h(r,\nu )=L(r,\nu)/M_{sh}$, acquired 
during expansion, integrating the torque over time (Ryden 1988a, Eq. 35): 
\begin{equation}
h(r,\nu ) = \frac{\tau_o t_o \overline{\delta}_o^{-5/2}}{\sqrt[3]{48}M_{sh}}%
\int_0^\pi \frac{\left( 1-\cos \theta \right) ^3}{%
\left( \vartheta -\sin \vartheta \right) ^{4/3}}\frac{f_2(\vartheta) \cdot%
d\vartheta}{f_1(\vartheta )-f_2(\vartheta )\frac{\delta _o}%
{\overline{\delta _o}}}
\label{eq:ang}
\end{equation}
The functions $f_1(\vartheta )$, $f_2(\vartheta )$ are given by Ryden (1988a) 
(Eq. 31), while the mean overdensity inside the shell, $\overline{\delta }%
(r)$, is given by Ryden (1988b):
\begin{equation}
\overline{\delta }(r,\nu )=\frac 3{r^3}\int_0^\infty d\sigma \sigma ^2\delta
(\sigma )
\label{eq:ma8}
\end{equation}
The mass dependence of the threshold parameter, $ \delta_{c}(\nu)$, can be
found as follows: we calculate the binding radius, $r_{b}$, of the shell 
using Hoffmann \& Shaham's criterion (1985):
\begin{equation}
T_{c}(r, \nu) \leq t_{0}
\label{eq:ma9}
\end{equation}
where $T_{c}(r,\nu)$ is the calculated time of collapse of a shell and $
t_{o}$ is the Hubble time. We find a relation between $ \nu$ and $M$ through
the equation $ M=4 \pi\rho r^{3}_{b}/3$. We so obtain $ \delta_{c}(\nu(M))$. 
In Fig. 1 we show the variation of the threshold parameter, $\delta _c(M)$,
with the mass $M$. Non-radial motions influence the value of $\delta _c$
increasing its value for peaks of low mass while
leaving its value unchanged for
high mass peaks. As a consequence, the
structure formation by low mass peaks is
inhibited. In other words, in agreement with the cooperative
galaxy formation theory (Bower 1993), structures form more
easily in over-populated regions. 
To get the temperature distribution it is necessary to know 
the temperature-mass
relation. This can be obtained using the virial theorem and energy
conservation (Bartlett \& Silk 1993). Adopting the same notation of 
Bartlett \& Silk we have that:
\begin{equation}
T=(6.4 keV)\left( \frac{M \cdot h}{10^{15}M_{\odot }}\right) ^{2/3}\left[ 1+%
\frac{ \sqrt[3]{2} \eta \psi \int \frac{L^2dr}{M^2r^3}}{(G^{2}%
 H_0^{2} \Omega_{0} M^{2})^{1/3}}\right] 
\label {eq:temp} 
\end{equation}
where $ \eta=r_{ta}/x_{1}$, being $r_{ta}$
the radius of the turn-around, while $x_{1}$ is defined by the relation
$M=4 \pi x^{3}_{1}\rho/3$ and $ \psi=r_{eff}/r_{ta}$ where $r_{eff}$
is the time-averaged radius of a mass shell.
Eq. (\ref{eq:temp}) was normalised to agree with Evrard's (1990) simulations
for $L=0$. \\
In Fig. 2 the temperature distribution derived taking into account
non-radial motions is compared with the X-ray temperature distribution 
predicted by a simple CDM model.
\begin{figure}[ht]
\psfig{file=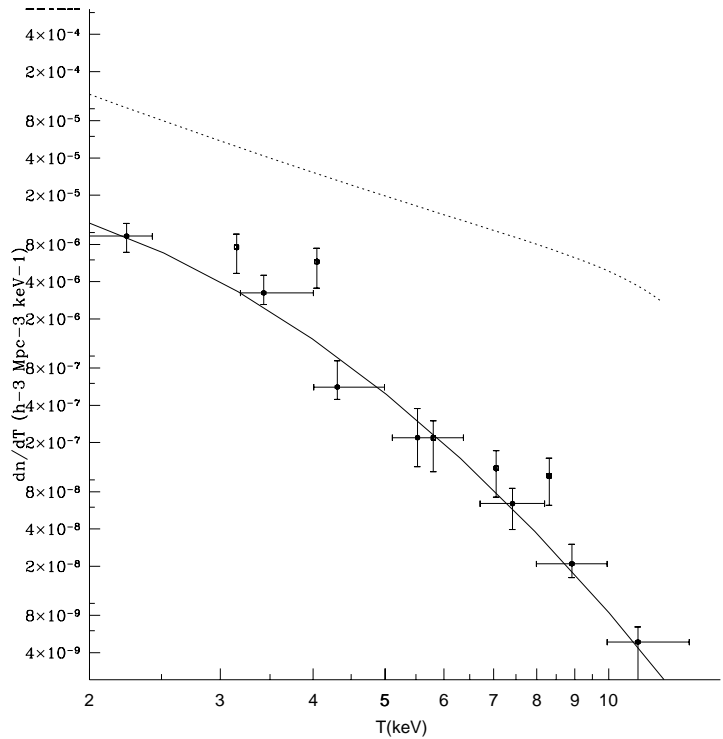,width=10cm}
\caption[]{X-ray temperature distribution function. The dashed line gives the
temperature function for a pure CDM model, with $ R_{f} = 3 Mpc$. The solid
line is the same distribution but now taking account of non-radial motions.}
\end{figure}
As shown the remarkable discrepancy  between pure CDM previsions and a CDM 
that taking account the non-radial motions is evident.
In a our paper (Del Popolo \& Gambera 1997) we compare our 
X-ray temperature distribution function with the experimental data 
obteined by Edge et al. (1990) and by Henry \& Arnaud (1991). \\

\section{Conclusion}

In this paper we have shown how non-radial motions may reduce the
discrepancy between the observed X-ray temperature distribution function of
clusters and that predicted by CDM model. As shown even with low
normalization, CDM power spectrum over-produces the cluster abundances. The
non-radial motions present in the outskirts of protoclusters have the effect
to reduce the cluster abundances producing a reasonable cluster temperature
distribution. Another feature of this model that shall be studied in a next
paper is the role of non-radial motions on the two-point correlation function
on scales larger than $10 Mpc$ (Gambera \& Del Popolo in preparation). 
We expect that our model shall 
be able to predict acceptable correlation functions.\\
 
\begin{flushleft}
{\it Acknowledgements}
\end{flushleft}
 We are grateful to E. Recami, E. Spedicato and to V. Antonuccio-Delogu for 
stimulating discussions during the period in which this work was performed.

\end{document}